\documentclass[a4paper]{aa}

\usepackage{psfig}
\usepackage{times}

\def\msun {\hbox{M$_{\odot}$}} 
 
\begin{document}
   \thesaurus{ 09.09.1; 09.19.2; 13.25.4}

    \title{The expansion of Cassiopeia A as seen in X-rays}

    \author{Jacco Vink \and Hans Bloemen \and Jelle S.  Kaastra \and 
Johan A. M. Bleeker}

    \offprints{J.Vink@sron.ruu.nl}

    \institute{SRON, Sorbonnelaan 2 NL-3584
              CA Utrecht, The Nether\-lands }

\date{Received  /  Accepted }

\maketitle

\begin{abstract}
We have for the first time measured the overall expansion rate of the 
supernova remnant Cas~A as seen in X-rays
using ROSAT and Einstein HRI images with time differences up to 
almost 17 years.
The overall expansion timescale of Cas~A is found to be $501\pm 15$~yr.
This is significantly shorter than
the timescale based on high resolution radio data.
Although the results clearly indicate that Cas~A is not anymore 
in the free expansion phase,
the discrepancy between the radio and X-ray timescales cannot be 
easily understood. Furthermore,
the expansion rate is incompatible with that predicted by a
self-similar model for a supernova remnant ploughing through the wind of 
its progenitor.
\end{abstract}

\section{Introduction}
The bright galactic radio source \object{Cassiopeia A} (Cas A) is believed 
to be the supernova remnant (SNR) of a massive star with a zero age main 
sequence mass in excess of 20\msun. The progenitor probably evolved into a 
Wolf-Rayet star type WN8 (Fesen et al. \cite{Fesen87}). 
Since Cas~A is also the youngest known galactic SNR, studying this
object potentially reveals important facts about the evolution, 
circumstellar medium and core collapse of its progenitor and, consequently, 
about massive stars in general.
Spectroscopic data (Chevalier \& Kirshner \cite{ChevK78}, \cite{ChevK79}) 
show that the 
remnant's blast-wave is ploughing through the helium and nitrogen enriched 
wind of the progenitor.
A recent estimate of the swept-up and ejected mass
based on the X-ray emissivity amounted to
$\sim 8\msun$\ and $\sim 4\msun$, respectively 	
(Vink et al. \cite{VKB96}).
The ratio of the two indicates that Cas~A is presumably not anymore in its 
free expansion phase, but rather in a transition phase between the
reverse-shock dominated phase and the adiabatic or Sedov phase of 
its evolution.

The best way to verify the dynamical status of Cas~A is to measure
the current expansion rate of Cas~A. 
This has been done in the optical (van den Bergh \& Kamper \cite{vdBergh83}) 
and in the radio 
(Tuffs \cite{Tuffs}, and Anderson \& Rudnick \cite{AR95} -- hereafter AR95 --) 
with different results. 
AR95 reported an expansion timescale of $864\pm 9$ yr for the radio knots and 
$750\pm 60$~yr for the diffuse component, 
whereas the optical expansion time indicates an explosion date of 
AD $1658 \pm 3$.
These results seem contradictory, but the optical knots consist of high density
gas ($10^3$~cm$^{-3}$) and suffer little deceleration. 
Furthermore, the optical knots comprise only a small ($< 1$\msun) 
fraction of the total mass of the remnant. 
So the optical expansion of Cas~A gives us
important information on the age of Cas~A, but little on the dynamical status.
The radio emission, on the other hand, is associated with the relativistic 
electrons and magnetic fields, rather than the bulk mass of Cas~A.
The most direct handle on the dynamical status of Cas~A comes from the 
emission closely associated with the bulk mass of Cas~A, i.e. 
the X-ray emission from the shock heated gas.

Some kinematical information based on the X-ray emission has already been 
provided by Doppler shift measurements of X-ray line complexes implying 
an expansion asymmetry between the Northwest and Southeast parts of the 
remnant with 
a velocity difference of $\sim$ 1500~km/s 
(Markert et al. \cite{Markert}, Holt et al. \cite{Holt}, Vink et al. \cite{VKB96}). 
The interpretation of the data is not clear: are we dealing here with
bipolar mass ejection, a ring-like morphology (Markert et al.) 
or small but measurable deviations from spherical symmetry? 
Vink et al. (\cite{VKB96}) reported a velocity line broadening of 
$\sim$4000~km/s, but it was later found that the ASCA SIS0, 
the instrument used to measure the broadening, 
suffered from a somewhat degraded spectral resolution.
So $\sim 4000$~km/s should be regarded as an upper limit to the actual 
line broadening (Vink et al. \cite{VKB97}).

Here we present a direct measurement of the angular expansion of Cas~A in 
X-rays over an interval of almost 17 years.

\section{Data and Method}\label{thedata}
\subsection{The data}
\begin{table}
	\caption{The observations used for this research.\label{data}}
	\begin{flushleft}
	\begin{tabular}{lllll}
        	\hline\noalign{\smallskip}
ID  &  Satellite & begin date & end date & exposure \\
 & & d/m/y & d/m/y & (ks) \\
        	\hline\noalign{\smallskip}
713       & Einstein &  08/02/1979  & 08/02/1979 & 42.5  \\
10139     & Einstein &  22/01/1981  & 23/01/1981 & 25.6  \\
150086    & ROSAT   &  29/07/1990  & 29/07/1990 & 8.0    \\
500444    & ROSAT   &  23/12/1995  & 01/02/1996 & 180.4  \\
        	\hline\noalign{\smallskip}
	\end{tabular}
	\end{flushleft}
\end{table}

\begin{figure}
	\psfig{figure=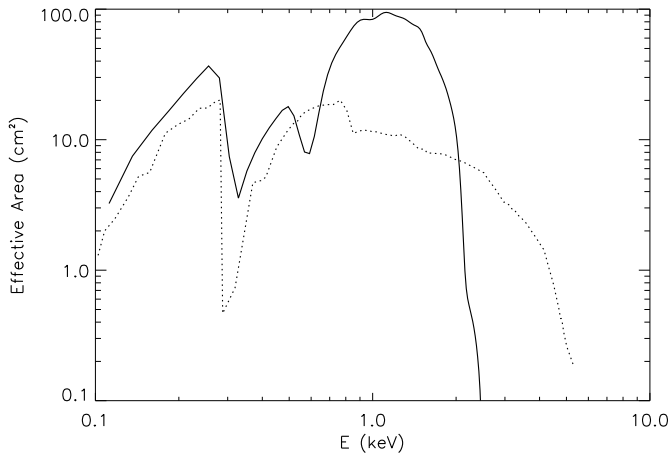,width=8.8cm}
	\caption[]{The ROSAT HRI (solid line) and the 
Einstein HRI (broken line) effective areas.\label{effArea}}
\end{figure}

Our measurement of the expansion of Cas~A is based on archival data 
of the ROSAT High Resolution Imager (RHRI) and the Einstein HRI
(EHRI). Table~\ref{data} gives an overview.
A detailed description of the RHRI and some information on the EHRI 
can be found in David et al. (\cite{David}). For the EHRI one can consult 
Giaccconi et al. (\cite{Giacconi}).
Both instruments are microchannel plate detectors and are similar in design; 
they were in fact built by the same hardware group.

The instruments have a spatial resolution of $\sim$4\arcsec\ FWHM, 
but the EHRI point spread function has broader wings. 
Note that the actual resolution of both instruments may be slightly worse
due to time dependent residual errors in the aspect solutions, 
which are of the order 1\arcsec.
The point spread function does not vary 
within off-axis angles of 5\arcmin, which is twice the radius of 
Cas~A. Consequently, off-axis effects are not important for our analysis.
The energy responses of the instruments are different as can be seen in
Fig.~\ref{effArea}.
The interstellar column density towards Cas~A is large, 
so most of the photons with energies below 1~keV are absorbed. 
This implies that most detected photons have energies between 1 and 2 keV,
but the EHRI image has an additional contribution from photons with 
energies up to $\sim$4~keV.
As for differences in plate scale, a study of M31 sources showed that 1 
RHRI pixel = $1.0050\pm 0.0007$ EHRI pixel, 1 RHRI pixel being 
$0.499\arcsec \pm 0.001\arcsec$ (David et al. \cite{David}).

The data we used were in the form of photon lists which have already 
been subjected to a basic reduction process. 
From these photon lists we made images which were converted to
the coordinate system of the 1995/96 RHRI image. 
After conversion the images were rebinned by a factor of 4 to a
nominal pixel size of 2\arcsec, 
so we oversample the resolution with a factor 2 to 3, 
compliant with Nyquist's criterion.

We corrected the long exposure of 1995/96 for the fact that it actually 
consists of many single exposures. 
The problem is that the attitude reconstruction of ROSAT is about 6\arcsec.
By matching the images of the individual exposures
we improved the quality of the combined image. 
However, because we only used exposures longer than 5~ks, 
our final image consists of 95~ks instead of 180~ks of data.
The rms spread in the attitude corrections was about 1.5\arcsec.
The procedure used for matching images was similar to the procedure for 
measuring expansion, which is described below.

\begin{figure*}[t]
\centerline{
	\psfig{figure=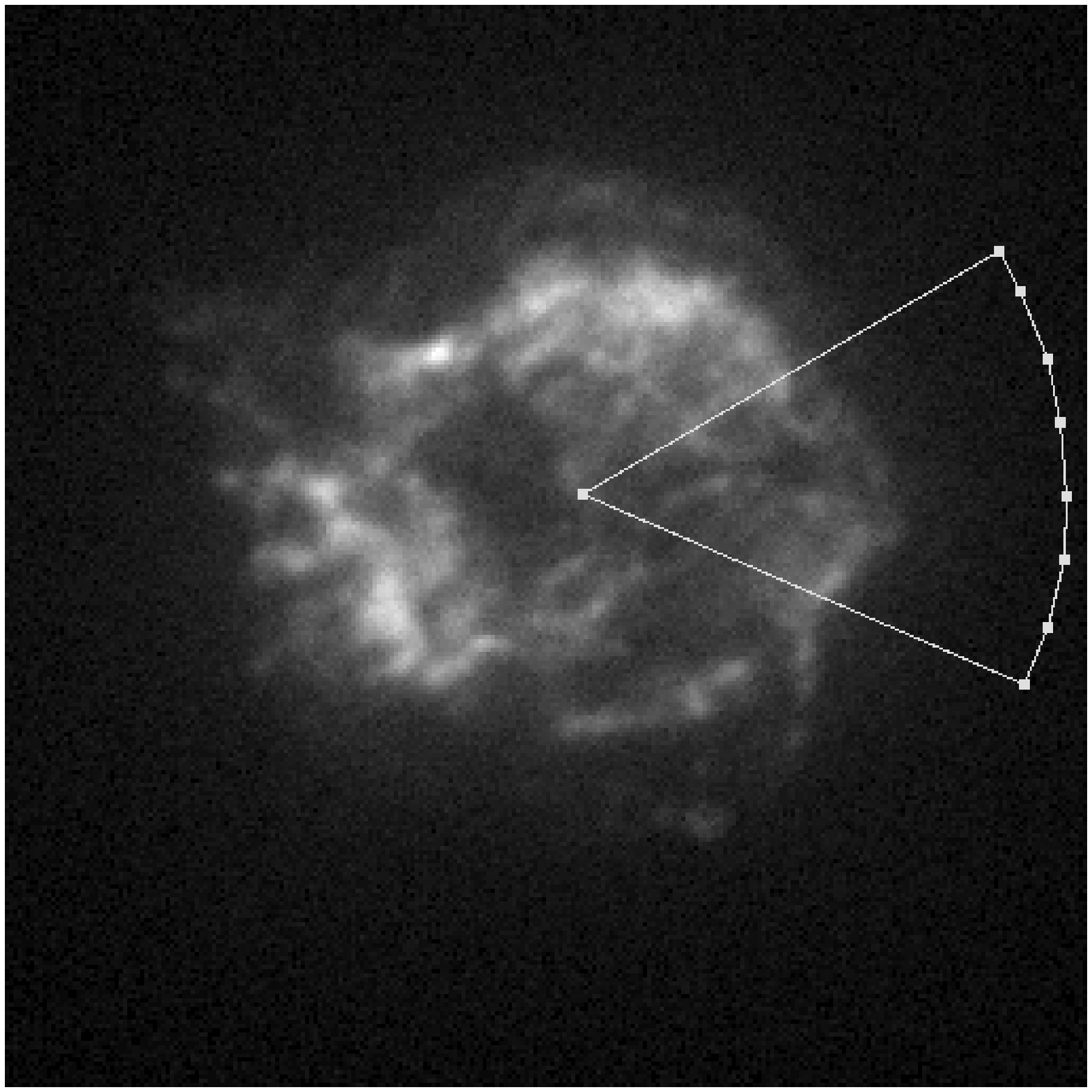,width=8.8cm}
	\psfig{figure=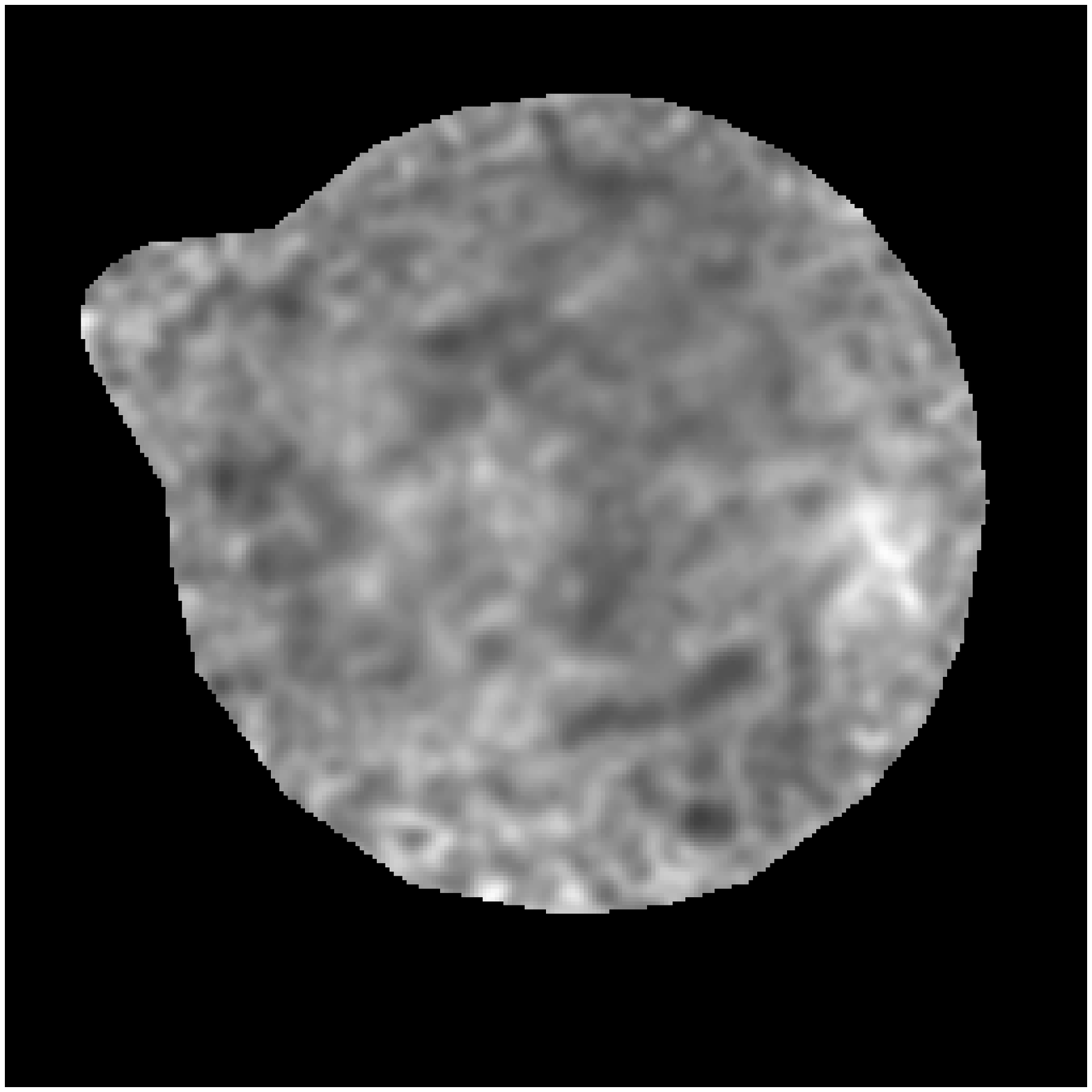,width=8.8cm}
}
	\caption[]{Left: 
The ROSAT HRI image (1995/96) in square root scaling with 
the region used for the expansion measurements of the Western region
superposed on it. The maximum of the image is 619 counts/pixel.
Right: the ratio image of the Einstein (1979) and the ROSAT 
(1995/96) image. The images were smoothed with a gaussian with 
$\sigma= 4\arcsec$ and the ROSAT image was corrected for expansion 
( so structures due to the expansion were removed).
The minimum ratio is 0.11 the maximum ratio is 0.43.
The region relevant for the expansion analysis is shown here.
The images are displayed on the same scale 
(the images are 8.5\arcmin\ by 8.5\arcmin).
\label{casA}}
\end{figure*}

\subsection{The method}
The expansion of Cas~A in X-rays can be qualitatively seen easily
by blinking the EHRI and the RHRI 1995/96 images
In fact, due to the differences in plate scale,
the actual expansion is even larger than viewed in this way. 
Our method of measuring the expansion is straightforward: 
we scale the latest image in such a way that no expansion
between the two images is discernible anymore.
Thereby we neglect small scale changes such as those occurring 
in individual knots.

Instead of judging by eye whether two images
match after rescaling one image, we used the maximum-likelihood method for a 
poissonian distribution (Cash \cite{Cash}) to see if the two images match.
The main advantage of the maximum-likelihood method over a $\chi^2$ fitting is
that the latter assumes that
the number of counts per pixel has a gaussian distribution, whereas
the maximum-likelihood method can be used for arbitrary distributions.
The maximum-likelihood method for a poissonian distribution
involves the maximization of the so called $C$ statistic.
The difference in the $C$ statistic 
(also called the log likelihood ratio) for different parameter values
has a $\chi^2$ distribution (Cash \cite{Cash}).

In all cases we use the RHRI 1995/96 image as the model image. 
We can neglect the statistical errors introduced by the model
image, since it is statistically far superior to any of the images used 
for comparison.
For example, the second best image from a statistical point of view is the 
Einstein image from 1979 which has a total number of photons a factor 7 less
than the model image. So the average statistical error per pixel is 
entirely dominated by the other images. 
Moreover, as we shall indicate below, 
the systematic errors dominate the statistical errors.
 
In addition to fitting an expansion factor, we also fitted the attitude 
correction,
which includes an image rotation, and an additional uniform background
level. 
The level of this uniform background can be determined empirically by 
introducing it as a free parameter, provided all pixel values of the model
image remain positive as required for the likelihood analysis.
As can be seen in Table~\ref{expansion} this condition was met in all three 
cases.
All corrections were only applied to the model image (the RHRI 1995/96 image); 
this image was rescaled, shifted and rotated using a method
described in Parker (\cite{Parker}):
each new pixel value is based on a linear interpolation of the 4 nearest 
pixels.
The model image was normalized in such a way that it contains the same number 
of counts as the images used for comparison.
In order to reduce poisson noise from the model image we smoothed 
the image with a gaussian filter with $\sigma = 1$\arcsec. 
This hardly influences the resolution of the model image.
The best fit expansion rate was barely influenced by the smoothing procedure.

In principle it is also possible to optimize for the
expansion center. However, there is a correlation between the expansion
center and the attitude correction, so they cannot be fitted independently.
For this reason the expansion center was fixed to the expansion center found 
by Reed et al. (\cite{Reed}) to be the best fit to the position/velocity 
distribution of optical knots 
( $\alpha_{2000}=23^h23^m26.6^s$, 
$\delta_{2000} = 58^\circ49\arcmin 1\arcsec$ ). 
We experimented with other expansion centers compiled
by Reed et al. (\cite{Reed}) and found that the expansion factors were very 
little affected by our choice of the actual expansion center 
(the expansion rate changed by less than $0.05\%$).
The expansion was evaluated within a given region of the image;
the region could be either a circle or a polygon.

The optimization of the likelihood was done by scanning the relevant
parameter space in increasingly smaller steps. By applying this method in 
an iterative way we tried to circumvent potential local maxima. 

\section{The expansion of Cas~A}
\begin{table*}
	\caption{
Results of our overall expansion measurements.
$f$ is the expansion between two images
not accounting for differences in plate scale, 
$\tau$ is the expansion timescale (epoch 1996),
$\delta x$ and $\delta y$ denote the attitude correction in pixels (2\arcsec) 
needed to match two images 
(positive values moves the model image towards the Southeast), 
$\phi$ denotes the rotational correction applied and the last column gives an 
additional background rate that was added to
the model image before normalizing the image to the image to which is was compared.
Only statistical errors are given. 
The systematic errors are of the order of 20~yr for the measurements involving
the Einstein HRI (i.e. $\Delta t > 14$~yr). 
See the text for a discussion of the estimated systematic errors.
\label{expansion}}
	\begin{flushleft}
	\begin{tabular}{llllllll}
        	\hline\noalign{\smallskip}
Dates&  $\Delta t$ &    $f$       &    $\tau$   & $\delta x$     & $\delta y$     &  $\phi$  & background\\
& (yr) & (\%) & (yr) & (pixel) & (pixel) &$(^\circ)$ & counts/pixel \\
        	\hline\noalign{\smallskip}
79 - 95/96 & 16.97 & $97.02\pm 0.04$ & $491\pm 6$  & 
  $0.86\pm 0.03$ & $0.22\pm 0.02$  & $-0.06\pm 0.02$ & $1.5\pm 0.1$ \\
81 - 95/96 & 14.98 & $97.42\pm 0.07$ & $489\pm 11$ &  
  $-0.14\pm 0.06$  & $0.80\pm 0.03$  & $-0.50\pm 0.04$ & $2.1 \pm 0.1$\\
90 - 95/96 & 5.44  & $98.94\pm 0.04$ & $513\pm 20$ & 
  $0.94\pm 0.03$ & $3.44\pm 0.03$  & $0.08\pm 0.04$ & $0.05 \pm 0.11$  \\
        	\hline\noalign{\smallskip}
	\end{tabular}
	\end{flushleft}
\end{table*}

With our method we measure the expansion factor, $f$, 
which is just the ratio of the sizes of two images.
The physically more interesting parameter is
the expansion timescale, $\tau$, which would be the age of the remnant 
in case no deceleration had taken place.
The relation between the two is:
\begin{equation}
\tau = {\Delta t \over{( 1 - s f)} },
\end{equation}
where $s$ denotes the plate scale ratio 
(0.995 in case an EHRI image is compared to a RHRI image) and $\Delta t$
is the time difference between the two exposures. As is clear from this
equation an error in $f$ weighs more heavily when $s f$ is close to 1.

Fig.~\ref{casA} (right) shows the region of the image
used for determining the expansion.
In Table~\ref{expansion} we list the values for $f$ and the expansion 
time scales for each time interval.
In case images of two different instruments are
compared we have to ensure that the different detector characteristic
(such as shown in Fig.~\ref{effArea}) do not affect the estimate of the
expansion timescale significantly.
We tested this using the following method.

If we know the expansion age of Cas~A we can, 
by correcting the Einstein image for the expansion,
estimate the differences of the response of the two detectors to the emission
of Cas~A. We do not know the exact expansion properties of Cas~A, but we do 
have the best fit values listed in Table~\ref{expansion}.
So by correcting for the expansion we can obtain a, first order, 
quantitative estimate of the effect of different detector properties on
the exposure of Cas~A.
We have visualized these differences by dividing
the combined Einstein images by the 1995/96 RHRI image on a pixel by
pixel basis.
In order to reduce the poisson noise we smoothed the images
with a gaussian with $\sigma = 4\arcsec$.
The resulting ratio image is shown in Fig.~\ref{casA} (right). 
Most of the structure seen in the ratio image is now allegedly arising from
differences in detector responses to the spectral characteristics of the X-ray
emission of Cas~A and to possible efficiency variations in the detectors.
The absence of ring-like structures in the ratio image indicates 
that we have corrected adequately for the expansion of Cas~A.
Note that the most salient feature in the ratio image, 
the high ratio in the West, can be attributed to the fact that 
the interstellar absorption peaks in the West of the remnant as OH 
and HI absorption studies indicate 
(Bieging \& Crutcher \cite{Bieging}, Schwarz et al. \cite{Schwarz}; 
see also Keohane et al. \cite{Keohane}).
 
To assess the potential sensitivity of our result to the differences displayed
in Fig.~\ref{casA}, we corrected the model image with these ratio's and
repeated the expansion measurements for this revised model.
Although some interdependence is now present in the image comparison,
this procedure is legitimate to bring out the influence if systematic 
effects due to relative differences in the instrumental response functions.
If we determine the expansion age of Cas~A using our revised
model, we find expansion rates that do not differ more than 10~yr 
from our best fit expansion rates. From this we conclude that the systematic 
error caused by different instrument characteristics is $\Delta f = 0.06\%$.
An additional systematic error is the uncertainty in the plate scale which is
$\Delta s = 0.07\%$.
Combining the two systematic errors we find that the total systematic error 
in the expansion age in case
an EHRI image is involved is of the order of $\Delta \tau = 19$~yr. 
This means that systematic errors dominate the statistical errors. 
From now on we will always include estimates of systematic errors.

Table~\ref{expansion} lists the individual expansion measurements.
Our overall best fit value for the expansion age of Cas~A is 
$\tau = 501 \pm 15$ (for the epoch 1996). This value is a weighted average
of the three uncorrected measurements, with the weight being the $1/\sigma^2$ 
value including a 19~yr systematic error for the expansion 
ages determined by a comparison between a ROSAT and an Einstein HRI image.
Note that, although the three measurements are not completely independent 
since they involve the 1995/96 image, our approach is again justified by the
superior statistical quality of the 1995/96 image.
The expansion age corresponds to an expansion rate of 
$0.200 \pm 0.006$~\%yr$^{-1}$. If we adopt a distance to Cas~A of 3.4~kpc
(Reed et al.~\cite{Reed}), the expansion age translates into a velocity 
of $\sim 3500$~km/s for the bright ring at a radius of 110\arcsec. 
The inferred velocity of the blast wave at a radius of about 160\arcsec\ 
is $\sim 5200$~km/s.

The expansion age of 501~yr is remarkable in the sense that it is 
significantly lower than the expansion age found in the radio (see AR95). 
The average expansion timescale of all radio knots was found to be 
$\tau = 864\pm 9$~yr, for the subsample of faint knots AR95 derived
$\tau = 747\pm 14$~yr (for the epoch 1987). 
An expansion age comparable to the latter value was derived
for the diffuse radio emission.
However, AR95  reported substantial variation of the expansion
timescale with azimuth. Their shortest timescales, i.e. the timescales
of the Eastern and Southeastern sectors, are consistent with our overall
expansion timescale. Furthermore, in X-rays the Western region is relatively
faint, whereas in the radio it is the brightest region. So there may be 
a bias in the radio measurements towards the Western region and a bias in our
measurement towards the expansion of the Eastern region. 

We also searched for variations in the expansion with azimuthal angle.
We found that, except for the Western region, all expansion timescales were 
consistent with the overall expansion timescale within the errors.
The sectors for doing the measurements were chosen to be similar to 
the sectors used by AR95, i.e. we used 6 sectors each spanning 
60$^\circ$.
For the Western region (sector V in AR95) we found 
$\tau = 734 \pm 144$ for $\Delta t = 5.44$ 
and a weighted average of $\tau = 620 \pm 51$ for the time intervals involving
the EHRI. In this case a correction for the instrumental efficiencies 
was applied.
All parameters except the expansion factor were 
fixed to the appropriate values listed in table~\ref{expansion}.
The expansion timescale is much larger for a comparison between the EHRI 
images and the 1995/96 image if we do not correct for the differences in 
instrumental efficiencies, 
namely a weighted average for measurements involving the 
EHRI of $\tau = 943\pm 134$ and a total weighted average of $\tau = 846\pm 98$.
So, unfortunately, the effect of the differences in detector characteristics
is largest for the Western region making the measurement 
unreliable (c.f. Fig.~\ref{casA} right). 
So we have an indication that the Western region 
is expanding more slowly than the rest of the remnant, 
although the evidence based on the X-ray emission alone is not conclusive.
A measurement of the overall expansion timescale of Cas~A excluding the
Western region gives $\tau = 467\pm 17$~yr.

\begin{figure}[td]
	\psfig{figure=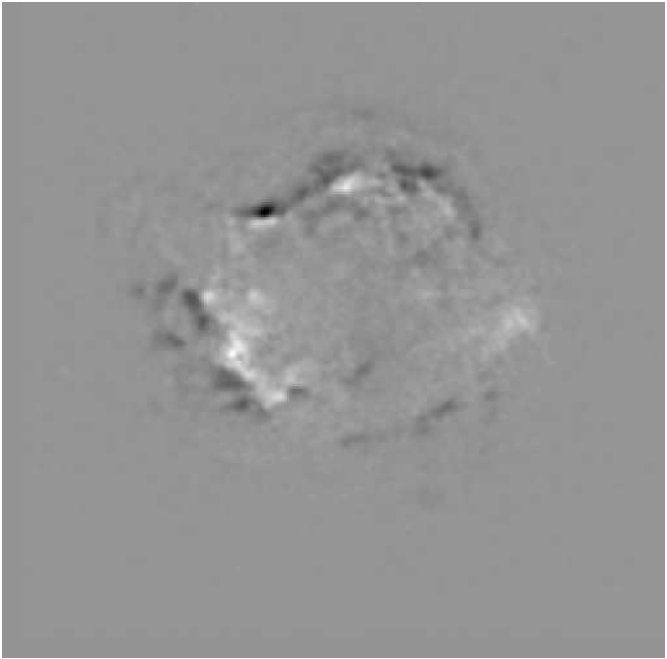,width=8.8cm}
	\caption[]{
A visualization of the expansion of Cas~A.
The ROSAT HRI image of 1995/96 was subtracted from the Einstein HRI image 
of 1979. The ROSAT image was normalized with respect to the Einstein image.
The ROSAT image has been corrected with the attitude correction from 
Table~\ref{expansion} and both images were smoothed with a gaussian with $\sigma = 2$\arcsec. Negative values are dark. The minimum of the image is -42 and the maximum is +29.\label{expim}}
\end{figure}

\section{The dynamics of Cas~A}
\subsection{The discrepancy between the X-ray and radio measurements}
Our measurement of the expansion age of Cas~A is significantly shorter than
the expansion age derived from proper motion measurements of compact radio 
features, although we should, as pointed out, beware of biases caused by the
different morphologies of the radio and X-ray image.
As put forward in the introduction, the advantage of an X-ray determination
of the expansion rate over a measurement based on the compact radio features is
that the bulk of the mass radiates in X-rays, whereas the compact radio 
features are associated with the magnetic fields and relativistic electrons.
This does, however, not mean that we can easily discard the radio data for 
that reason, as the compact radio features, 
some of which show remarkable resemblance to bow shocks, can be comprehensively
described as ejecta with a higher than average density and moving 
supersonically through diffuse X-ray emitting gas 
(Anderson et al. \cite{Anderson}, AR95). 
This means that it is expected that the diffuse gas is moving more slowly than 
these compact features.
Clearly, our measurement is at odds with this interpretation.
However, the kinematics of the Western region in the radio 
band do display phenomena that are not easily explained by a simple dynamical
model. For instance, the compact radio features in the Western region 
show large deviations from a radially outward motion. 
Some features even display an inward motion (Anderson et al. \cite{Anderson}).
Such deviations do bias the expansion rate of the radio features towards a
longer timescale, but at the moment we lack a clear understanding of this 
phenomenon. Another peculiarity which is hard to understand from the point of 
view of the overall dynamics of Cas~A is a small but significant net 
acceleration of the radio features towards the North.
So our understanding of the compact radio features is still 
incomplete. 
The discrepancy between our expansion rate and 
the expansion rate based on the compact radio features certainly deserves 
serious future attention.
We therefore concentrate on the implications of the shorter expansion 
timescale for the dynamics of Cas~A derived from the analysis presented here.

\subsection{The reverse-shock model}
The explosion leading to Cas~A probably took place in 1680, that is if one
assumes that the 6th magnitude star observed by Flamsteed was indeed 
the supernova (Ashworth 1980). 
Although there are some problems with this identification,
this explosion date is supported by the kinematics of the subset of fast 
moving optical knots which are probably least decelerated (Fesen et al. 1987). 
So we assume that in 1996, the year of the last RHRI observation of Cas~A, 
the remnant was 316 yr old. 
Combining this with our overall expansion timescale of 501~yr we arrive at
a deceleration parameter of $m = 0.63\pm 0.02$ 
(the ratio of the true age over the expansion age). 
The deceleration parameter equals the ratio of the current velocity over the 
average velocity. 
To put this value in a simple theoretical framework we refer to the 
self-similar hydrodynamical model of Chevalier (\cite{Chev82}). 
The Chevalier model includes a blast-wave and a reverse-shock. 
The model assumes that the density profile of the unshocked gas is distributed
as $\rho \sim r^{-n}$\ for the ejecta component and $\rho \sim r^{-s}$ for 
the circumstellar medium with $n$ and $s$ as free parameters.
Optical spectroscopy (Chevalier \& Kirshner \cite{ChevK78}, \cite{ChevK79}) 
indicates that Cas~A is expanding into the wind of its progenitor, 
which implies $s=2$. For type II supernovae the 
density distribution of ejecta is probably rather steep, $n\sim 10$, 
but for more compact stars, 
such as the Wolf-Rayet star that was the progenitor of Cas~A,
the density distribution may be less steep.

The radius of the
contact discontinuity between ejecta and circumstellar matter evolves in the
Chevalier model as $R_c \propto t^{{n-3}\over{n-s}}$. 
The case $n=5$ corresponds to the Sedov (\cite{Sedov}) evolution 
and $n < 5$ does not correspond to any physical solution. 
The deceleration parameter for the Chevalier model is $m= {n-3\over{n-s}}$.
So for $s=2$ and the observed value of $m = 0.63$ we need $n = 4.70\pm 0.15$ 
which is very close to the Sedov solution. However, if taken
literally it would mean that Cas~A is blast-wave instead of 
reverse-shock dominated as generally believed. 
One can of course assume that $s=2$ is a reasonable but not quite good 
representation of the structure of the circumstellar medium around the
progenitor. For example, $s=1$ gives $n=6.4$, which gives a valid Chevalier
model, but lacks the conceptual simplicity of a $s=2$ model.
Another assumption of the Chevalier model that may not be valid is 
the power law density distribution  of the ejecta.
Only the outer layers of ejecta are thought to have a distribution
well represented by a power law. 
If, however, a substantial fraction of the ejecta has already been shocked,
then the evolution of Cas~A has entered a phase for which the Chevalier model 
is not applicable anymore. The fact that for $s=2$ we arrive
at a solution close to the Sedov model may indeed indicate that Cas~A is
in a transition phase from a reverse-shock dominated to a blast-wave dominated
(Sedov) phase.
A model in which most of the ejecta have already been 
shocked is in agreement with the mass ratio of the swept up and ejected 
mass proposed by Vink et al. (1996).
Note that also the X-ray spectrum of Cas~A cannot be well described by
a $s=2$ Chevalier model because the model implies a much larger ratio 
between the reverse-shock and blast-wave temperatures than observed
(Jansen et al. 1988, Vink et al. 1996). A $s=0 $ Chevalier model provides
a better fit to the X-ray spectrum, but has as the drawback that we know
that Cas~A is moving through the wind of its progenitor.

Apart from the ejecta density distribution the circumstellar medium
may also be more complicated than assumed by the Chevalier models.
Indeed, two recent numerical models indicate that the the circumstellar
medium may include a dense shell of material originating in the
red supergiant phase of the progenitor and swept up by the fast wind
of the Wolf-Rayet star
(Garc\'\i a-Segura et al. \cite{GLM96} and Borkowski et al. \cite{Bork96}).
The study by Borkowski et al. (\cite{Bork96}) was made in order to reproduce
the expansion rate of Cas~A as observed in the radio (AR95) 
which, however, may not be the correct value as this study indicates.
The possible existence and position of such a shell 
(it may also reside outside the current blast-wave) can provide valuable
information about the detailed history and properties of
the progenitor, as shown by Garc\'\i a-Segura et al. (\cite{GLM96}).
However, the numerical models are too specific to see how our result would 
change their conclusions.

\subsection{An encounter with a molecular cloud?}
Both numerical models do not pay attention to inhomogeneities that may 
have led to a possible slower expansion of the Western region or
to the differences in expansion velocities of the optical knots 
between the front and the back of the remnant (Reed et al. \cite{Reed}). 
An explosion in an off-center bubble is a possibility 
(Reed et al. \cite{Reed}).
The asymmetry of the bubble may be caused by existing inhomogeneities in the 
interstellar medium. Another interesting and possibly related suggestion 
(Keohane et al. \cite{Keohane}, AR95) 
is that in the West Cas~A is interacting with a molecular cloud seen in OH 
absorption towards Cas~A (Bieging \& Crutcher \cite{Bieging}).
This is the same cloud that is probably responsible for the relative hardness
of the Western region seen in Fig.~\ref{casA}.
In the case of an interaction with a molecular cloud it is expected
that the Western region would be relatively bright.
This is in fact the case when the X-ray emission is corrected for 
the absorption towards Cas~A (Keohane et al. \cite{Keohane})
which varies strongly over the remnant.

\section{Conclusion}
This is the first study of the expansion rate of Cas~A as seen in X-rays. 
A similar study was done for \object{Tycho}'s SNR 
(Hughes \cite{Hughes97}) using a 
somewhat different method. Measuring the expansion timescale
of a SNR in X-rays has one major advantage over similar studies in the radio
and the optical bands: 
the X-ray emission is directly connected  to the bulk mass of SNRs. 

We have found an overall expansion timescale of $501 \pm 15$~yr corresponding 
to an expansion rate of $0.200 \pm 0.006$~\%yr$^{-1}$. 
The optical expansion timescale is shorter, namely of the order of 340~yr.
This can be easily understood, since the optical knots consist of dense
gas which is expected to be less decelerated than the bulk of the ejecta.
More intricate is the discrepancy between our expansion timescale and the
expansion timescale based on the compact radio features ($\sim800$~yr),
since also these radio features are thought to be the marks of denser
gas moving through a more diffuse medium.
This explanation is clearly not in agreement with our expansion timescale 
as we discussed in the previous paragraph. 
However, we lack a coherent picture of the dynamics of
the compact radio features. Furthermore, as we already pointed out, 
the radio image is brightest in the West of Cas~A, whereas at the 
photon energies where the ROSAT and Einstein HRI are most sensitive Cas~A is 
brightest in the East. This may have caused different biases
in the radio and X-ray measurements.

An expansion timescale of 501~yr means that the standard reverse-shock
Chevalier (\cite{Chev82}) model for a SNR moving through the wind of its 
progenitor is
not applicable to Cas~A. It either means that Cas~A is in a transition
phase between the reverse-shock dominated and the blast-wave dominated phase
or it means that the situation is too complex to be well
described by simple analytical SNR models.

In the near future high spatial resolution images of Cas~A obtained with AXAF
will be able to improve the existing measurements.
Meanwhile there are other outstanding issues concerning the dynamics 
of Cas~A such as the front back asymmetry of optical knots 
(Reed et al. \cite{Reed}) 
and the North/South asymmetry in the Doppler shifts observed in 
the X-ray spectra (Markert et al. \cite{Markert}). 
Together with the expansion rate these facts
should ultimately be incorporated into a complete three dimensional dynamical
model of Cas~A.

{\it Note.} During the refereeing process a preprint appeared with 
a similar measurement as presented in this paper 
(Koralesky et al.~\cite{Koralesky}). Their results confirm our analysis.

\begin{acknowledgements}
We thank Lucien Kuiper, Frits Paerels and Karel van de Hucht for useful 
information and discussions.
This research has made use of data obtained through the High Energy
Astrophysics Science Archive Research Center Online Service, provided
by the NASA/Goddard Space Flight Center and the ROSAT Archive at the 
Max Planck Institut f\"ur Extraterrestische Physik.
This work was financially supported by NWO, 
the Netherlands Organization for Scientific Research.
\end{acknowledgements}


\begin{thebibliography}{}
	\bibitem[1994]{Anderson} Anderson M.C., Jones T.W., Rudnick L., et al., 1994, ApJ 421, L31
	\bibitem[1995]{AR95} Anderson M.C., Rudnick L., 1995, ApJ 441, 307 (AR95)
	\bibitem[1980]{Ashworth} Ashworth W.B., 1980, J. Hist. Astr., 11, 1
	\bibitem[1986]{Bieging} Bieging J.H., Crutcher R.M., 1986, ApJ 310, 853
	\bibitem[1996]{Bork96} Borkowski K., Zymkowiak A.E., Blondin J.M., Sarazin C. L., 1996, ApJ 466, 866
	\bibitem[1979]{Cash} Cash W., 1979, ApJ 228, 939
	\bibitem[1982]{Chev82} Chevalier R.A., 1982, ApJ 258, 790
	\bibitem[1978]{ChevK78} Chevalier R.A., Kirshner R.P., 1978, ApJ 219, 931
	\bibitem[1979]{ChevK79} Chevalier R.A., Kirshner R.P., 1979, ApJ 233, 154
	\bibitem[1997]{David} David L.P., Harnden F.R., Kearns K.E., et al., 1997, ``The ROSAT HRI Calibration Report'', U.S. ROSAT Science data center/SAO, Harvard USA 
	\bibitem[1987]{Fesen87} Fesen R.A., Becker R.H., Blair W.P., 1987, ApJ 313, 378
	\bibitem[1996]{GLM96} Garc\'\i a-Segura G., Langer N., Mac Low M.-M., 1996, A\&A 316, 133
	\bibitem[1979]{Giacconi} Giacconi, R., Branduardi, G., Briel, U., et al., 1979, ApJ 230, 540
	\bibitem[1994]{Holt} Holt S.S., Gotthelf E.V., Tsunemi H., Negoro H., 1994, PASJ 46, L15
	\bibitem[1997]{Hughes97} Hughes J.,  1997, in ``X-ray Imaging and Spectroscopy of Cosmic Hot Plasmas'', Eds. F.Makino and K. Mitsuda, Universal Academy Press, Tokyo.
	\bibitem[1988]{Jansen} Jansen F.A., Smith A., Bleeker J.A.M., et al., 1988, ApJ 331, 9
	\bibitem[1998]{Koralesky} Koralesky B., Rudnick L., Gotthelf E.V., 
Keohane J.W., 1998, submitted to ApJL
	\bibitem[1996]{Keohane} Keohane J.W., Rudnick L., Anderson M.C., 1996, ApJ 466, 309
	\bibitem[1983]{Markert} Markert T.H., Canizares C.R., Clark G.W., Winkler P.F., 1983, ApJ 268, 13
	\bibitem[1994]{Parker} Parker J.R., 1994, ``Practical Computer Vision using C'', John Wiley \& Sons, Inc., New York
	\bibitem[1995]{Reed} Reed J.E., Hester J.J., Fabian A.C, Winkler P.F., 1995, ApJ 440, 706
	\bibitem[1959]{Sedov} Sedov L., 1959, ``Similarity and Dimensional Methods in Mechanics'', New York, Academic Press
	\bibitem[1997]{Schwarz} Schwarz U.J., Goss W.M., Kalberla P.M.W., 1997, A\&AS 123, 43
	\bibitem[1986]{Tuffs} Tuffs R.J., 1986, MNRAS 219, 13
	\bibitem[1983]{vdBergh83} Van den Bergh S., Kamper K. W., 1983, ApJ 268, 129
	\bibitem[1996]{VKB96} Vink J., Kaastra J.S., Bleeker J.A.M., 1996, A\&A 307, L41
	\bibitem[1997]{VKB97} Vink J., Kaastra J.S., Bleeker J.A.M., 1997, in ``X-ray Imaging and Spectroscopy of Cosmic Hot Plasmas'', Eds. F.Makino and K. Mitsuda, Universal Academy Press, Tokyo.
\end{thebibliography}
\end{document}